\documentclass[prb,showpacs,showkeys]{revtex4}

\usepackage{graphicx}
\usepackage{dcolumn}
\usepackage{bm}

\begin{document}

\title{Dirac quantization of a nonminimal gauged O(3) sigma model}%
\author{K. C. Mendes}

\affiliation{N\'ucleo de F\'{\i}sica, Universidade Estadual do
Cear\'a, Av. Paranjana, 1700, CEP 60740-000, Fortaleza, Cear\'a,
Brazil\footnote{Present address} \\}  \affiliation{Departamento de
F\'{\i}sica, Universidade Federal do Cear\'a, Caixa Postal 6030,
60455-760, Fortaleza, Cear\'a, Brazil
\\ }

\author{R. R. Landim}%
 \affiliation{Departamento de
F\'{\i}sica, Universidade Federal do
Cear\'a, Caixa Postal 6030, 60455-760, Fortaleza, Cear\'a, Brazil}%
\author{C. A. S. Almeida}
\email{carlos@fisica.ufc.br} \affiliation{Departamento de
F\'{\i}sica, Universidade Federal do
Cear\'a, Caixa Postal 6030, 60455-760, Fortaleza, Cear\'a, Brazil}%

\begin{abstract}
The (2+1) dimensional gauged O(3) nonlinear sigma model with
Chern-Simons term is canonically quantized. Furthermore, we study a
nonminimal coupling in this model implemented by means of a
Pauli-type term. It is shown that the set of constraints of the
model is modified by the introduction of the Pauli coupling.
Moreover, we found that the quantum commutator relations in the
nominimal case is independent of the Chern-Simons coefficient, in
contrast to the minimal one.
\end{abstract}


\maketitle

\section{Introduction}
The importance of O(3) nonlinear sigma model lies in theoretical and
phenomenological basis. It is a fact that this theory describe
classical (anti) ferromagnetic spin systems at their critical points
in Euclidean space, while in the Minkowski one it delineates the
long wavelength limit of quantum antiferromagnets. The model
exhibits solitons, Hopf instantons and novel spin and statistics in
2+1 space-time dimensions with inclusion of the Chern-Simons term.
The gauging of a nonlinear sigma model, given rise to a coupling
between the scalar fields (coordinates of the target space) and the
gauge fields has attracted interest from different areas. In
particular soliton solutions of the gauged O(3) Chern-Simons model
may be relevant in planar condensed matter systems
\cite{zee,lauglin,polyakov}. Recently gauged gravitating nonlinear
sigma model was considered in order to obtain self-dual cosmic
string solutions \cite{larsen}.

The canonical quantization of nonlinear gauged sigma models has been
studied by some authors in the context of the $CP^{1}$ sigma model.
Indeed this question was considered in connection with fractional
spin by Panigrahi and collaborators \cite{pani,pani1} and later by
Han \cite{han}. On the other hand, the canonical quantization of the
O(3) nonlinear sigma model with the Hopf term was treated by Bowick
{\it et al.} \cite{bowick} in order to compute the angular momentum
and establish the relation of a fractional spin and the coefficient
of the Hopf term.

In this work we investigate the soliton sector of the O(3) nonlinear
sigma model coupled to an Abelian gauge field through a nonminimal
term and in the presence of the Chern-Simons term \cite{marcon},
using the Dirac formalism for constrained systems \cite{dirac}. The
nonminimal term could be interpreted as a generalization of the
Pauli coupling, {\it i. e. }, an anomalous magnetic moment. It is a
specific feature of (2+1) dimensions that the Pauli coupling exists
not only for spinning particles, but for scalars ones too
\cite{kogan,lat,torres,georgelin}.

\section{The minimal gauged O(3)
nonlinear sigma model} \label{sec:1}  Let us begin by considering
the {\bf minimal} gauged O(3) nonlinear sigma model in a covariant
gauge-fixed form. Models with an explicit gauge dependence may be
suitable; for instance $\alpha$ in the Nakanishi-Lautrup formalism:
\begin{equation}
L=\frac{1}{2}D_{\mu }{\bm \phi }\cdot D^{\mu }{\bm \phi  }+\frac{\kappa }{2}%
\epsilon ^{\mu \nu \lambda }A_{\mu }\partial _{\nu }A_{\lambda
}-A_{\mu }\partial ^{\mu }b+\frac{\alpha }{2}b^{2}, \label{kcap3e1}
\end{equation}
where
\begin{equation}
\ D_{\mu }{\bm \phi} ={\partial }_{\mu }{\bm \phi +}eA_{\mu }{\bf
n\times \bm \phi}, \label{kcap3e2}
\end{equation}
is the minimal gauge covariant derivative and the second term is the
Abelian Chern-Simons term. The addition of this term to the O(3)
sigma model makes the vector field propagate even at the classical
level. The multiplier field $b$ is introduced in order to implement
the Lorentz covariant gauge-fixed condition. The canonical momenta
conjugate to the fields \ $A_{0},$ $A_{k},$ $b,$
$\phi _{1},$ $\phi _{2},$ $\phi _{3}$ respectively, are given by ($%
i,j,k=1,2$)
\begin{equation}
\pi _{0}=0,  \label{kcap3e4}
\end{equation}
\begin{equation}
\pi ^{k}=\frac{\kappa }{2}\epsilon ^{ki}A_{i},  \label{kcap3e5}
\end{equation}
\begin{equation}
\pi _{b}=-A_{0},  \label{kcap3e6}
\end{equation}
\begin{equation}
\Pi _{1}=\dot{\phi}_{1}-eA_{0}\phi _{2},  \label{kcap3e7}
\end{equation}
\begin{equation}
\Pi _{2}=\dot{\phi}_{2}+eA_{0}\phi _{1},  \label{kcap3e8}
\end{equation}
\begin{equation}
\Pi _{3}=\dot{\phi}_{3}.  \label{kcap3e9}
\end{equation}

Following the Dirac procedure for constrained systems,
non-derivative canonically conjugate momenta are classified as
primary constraints. Therefore, our primary constraints set is
\begin{equation}
V_{1}=\pi _{0}\approx 0,  \label{kcap3e10}
\end{equation}
\begin{equation}
V_{2}={\bm \phi }\cdot {\bm \phi }-1\approx 0 \label{kcap3e11}
\end{equation}
\begin{equation}
V_{3}=\pi _{1}+\frac{\kappa }{2}A_{2}\approx 0,  \label{kcap3e12}
\end{equation}
\begin{equation}
V_{4}=\pi _{2}-\frac{\kappa }{2}A_{1}\approx 0,  \label{kcap3e13}
\end{equation}
\begin{equation}
V_{5}=\pi _{b}+A_{0}\approx 0.  \label{kcap3e14}
\end{equation}

So these constraints leads us to the canonical Hamiltonian
\begin{equation}
H_{c}=\pi _{0}\dot{A}_{0}+\pi ^{k}\dot{A}_{k}+\pi _{b}\dot{b}+\Pi _{j}\dot{%
\phi}_{j}-L  ,\label{kcap3e15}
\end{equation}
which can be put in the form
\begin{eqnarray}
H_{C} &=&-\frac{1}{2}D^{j}{\bm \phi }\cdot D_{j}{\bm \phi -}\frac{{\bf \Pi }%
\cdot {\bf \Pi }}{2}+eA_{0}\left( \phi _{2}\Pi _{1}-\phi _{1}\Pi
_{2}\right)
\nonumber \\
&&-\kappa\epsilon ^{ij}A_{0}\partial _{i}A_{j}+A_{j}\partial ^{j}b-\frac{\alpha }{%
2}b^{2} . \label{kcap3e16}
\end{eqnarray}

Preservation of the primary constraints (\ref{kcap3e10}),
(\ref{kcap3e11}), (\ref{kcap3e12}), (\ref{kcap3e13}) and
(\ref{kcap3e14}), yield the consistent condition
\[
\dot{V}_{j}\approx 0    \qquad.
\]

Now is it necessary to implement the primary constraints in the
theory. In order to do this, we introduce a set of Lagrange
multipliers $\lambda _{1},\lambda _{2},\lambda _{3},\lambda _{4}$
and $\lambda _{5}$ in the canonical Hamiltonian. So we arrive in the
so called Dirac Hamiltonian
\begin{equation}
H=H_{c}+(\lambda _{1}V_{1}+\lambda _{2}V_{2}+\lambda
_{3}V_{3}+\lambda _{4}V_{4}+\lambda _{5}V_{5}).  \label{kcap3e17}
\end{equation}
The fundamental Poisson brackets are
\begin{equation}
\lbrack A_{0},\pi _{0}]=\delta (x-y) ,  \qquad     \lbrack A_{i},\pi
_{j}]=\delta _{ij}\delta (x-y),  \label{kcap3e18}
\end{equation}
and
\begin{equation}
\lbrack \phi _{i},\Pi _{j}]=\delta _{ij}\delta (x-y), \qquad \lbrack
b,\pi _{b}]=\delta (x-y) ,\label{kcap3e20}
\end{equation}
the others being zero. The time evolution of the primary constraints
with the Dirac Hamiltonian (\ref{kcap3e17}) determines all the
multiplier fields. Namely,
\begin{equation}
\dot{V}_{1}=[V_{1},\int d^{2}xH]=[\pi _{0},\int d^{2}xH_{c}]+\lambda
_{5}\approx 0,  \label{kcap3e22}
\end{equation}
and
\begin{equation}
\dot{V}_{2}=[V_{2},\int d^{2}xH]=[{\bm \phi }\cdot {\bm \phi
}-1,\int d^{2}xH_{c}]={\bf \Pi }\cdot {\bm \phi }\approx 0.
\label{kcap3e23}
\end{equation}

There is no more secondary constraints engendered by the consistency
condition for $V_{3}$ e $V_{4}$, since $[V_{3},V_{4}]\neq 0$. The
Lagrange multiplier $\lambda _{5}$ is obtained from
(\ref{kcap3e22}). On the other hand, the constraint $V_{6}={\bf \Pi
}\cdot {\bm \phi }\approx 0$ will give us more one secondary
constraint, namely
\begin{equation}
\dot{V}_{6}=[V_{6},\int d^{2}xH]=[{\bf \Pi }\cdot {\bm \phi },\int
d^{2}xH_{c}]+\lambda _{2}\approx 0 .  \label{kcap3e24}
\end{equation}
However the consistency condition will determine the Lagrange
multiplier $\lambda _{2}$, and we will have not any further
secondary constraints in the theory. Therefore our set of fully
second-class constraints are
\begin{equation}
V_{1}=\pi _{0}\approx 0 , \label{kcap3e25}
\end{equation}
\begin{equation}
V_{2}={\bm \phi }\cdot {\bm \phi }-1\approx 0 , \label{kcap3e26}
\end{equation}
\begin{equation}
V_{3}=\pi _{1}+\frac{\kappa }{2}A_{2}\approx 0 ,  \label{kcap3e27}
\end{equation}
\begin{equation}
V_{4}=\pi _{2}-\frac{\kappa }{2}A_{1}\approx 0 ,  \label{kcap3e28}
\end{equation}
\begin{equation}
V_{5}=\pi _{b}+A_{0}\approx 0 ,  \label{kcap3e29}
\end{equation}
\begin{equation}
V_{6}={\bm \phi \cdot \bf \Pi }\approx 0 .  \label{kcap3e30}
\end{equation}

Note that, there is not first-class constraints in the theory. This
fact is due we choose to work with a theory in a covariant
gauge-fixed form. As is it known the second-class constraints must
be eliminated since they do not generate physical transformations.
In order to do this we introduce the Dirac brackets defined as
\begin{equation}
\lbrack \Theta (x),\theta (y)]_{D}=[\Theta (x),\theta (y)]-[\Theta
(x),V_{i}(\zeta )](C^{-1})_{ij}[V_{j}(\xi ),\theta (y)]
\label{kcap3e34}
\end{equation}
where $C_{ij}(x,y)$ is an invertible matrix defined by
\begin{equation}
C_{ij}=[V_{i}(x),V_{j}(y)] . \label{kcap3e31}
\end{equation}
The $C_{ij}$ in this case has the form
\begin{equation}
C(x,y)=\left[
\begin{array}{cccccc}
0 & 0 & 0 & 0 & -1 & 0 \\
0 & 0 & 0 & 0 & 0 & 2 \\
0 & 0 & 0 & \kappa & 0 & 0 \\
0 & 0 & -\kappa & 0 & 0 & 0 \\
1 & 0 & 0 & 0 & 0 & 0 \\
0 & -2 & 0 & 0 & 0 & 0
\end{array}
\right] \delta (x-y).  \label{kcap3e32}
\end{equation}
and its inverse is written as
\begin{equation}
(C)^{-1}(x,y)=\left[
\begin{array}{cccccc}
0 & 0 & 0 & 0 & 1 & 0 \\
0 & 0 & 0 & 0 & 0 & -\frac{1}{2} \\
0 & 0 & 0 & -\frac{1}{\kappa} & 0 & 0 \\
0 & 0 & -\frac{1}{\kappa} & 0 & 0 & 0 \\
-1 & 0 & 0 & 0 & 0 & 0 \\
0 & \frac{1}{2} & 0 & 0 & 0 & 0
\end{array}
\right] \delta ^{-1}(x-y).  \label{kcap3e33}
\end{equation}

Therefore the set of nonvanishing Dirac brackets is given below
\begin{equation}
\lbrack A_{0},b]_{D}=\delta (x-y)  \label{kcap3e35}
\end{equation}
\begin{equation}
\lbrack A_{i},\pi _{j}]_{D}=\frac{1}{2}\delta _{ij}\delta (x-y)
\label{kcap3e36}
\end{equation}
\begin{equation}
\lbrack A_{i},A_{j}]_{D}=\frac{\varepsilon _{ji}}{\kappa}\delta
(x-y) \label{kcap3e37}
\end{equation}
\begin{equation}
\lbrack \pi _{i},\pi _{j}]_{D}=\frac{\kappa}{4}\varepsilon
_{ji}\delta (x-y) \label{kcap3e38}
\end{equation}
\begin{equation}
\lbrack \phi _{i},\Pi _{j}]_{D}=\left( \delta _{ij}-\phi _{i}(x)\phi
_{j}(y)\right) \delta (x-y) \label{kcap3e39}
\end{equation}
\begin{equation}
\lbrack \Pi _{i},\Pi _{j}]_{D}=\left( \phi _{j}(y)\Pi _{i}(x)-\phi
_{i}(x)\Pi _{j}(y)\right) \delta (x-y) \label{kcap3e40}
\end{equation}

Quantization follows in the usual way by replacing $\ i[F,G,]_{D}\longrightarrow \left[ \hat{F},\hat{G}%
\right] $, where $ \hat{F}$  and $\hat{G}$  denote operators. Then
we obtain
\begin{equation}
\lbrack \hat{A}_{0},\hat{b}]=i\delta (x-y)  \label{kcap3e41}
\end{equation}
\begin{equation}
\lbrack \hat{A}_{i},\hat{\pi}_{j}]=\frac{i}{2}\delta _{ij}\delta
(x-y) \label{kcap3e42}
\end{equation}
\begin{equation}
\lbrack \hat{A}_{i},\hat{A}_{j}]=\frac{i}{\kappa}\varepsilon
_{ji}\delta (x-y) \label{kcap3e43}
\end{equation}
\begin{equation}
\lbrack \hat{\pi}_{i},\hat{\pi}_{j}]=\frac{i\kappa}{4}\varepsilon
_{ji}\delta (x-y)  \label{kcap3e44}
\end{equation}
\begin{equation}
\lbrack \hat{\phi}_{i},\hat{\Pi}_{j}]=i\left( \delta _{ij}-{\hat{\phi}%
_{i}(x)\hat{\phi}_{j}(y)}\right) \delta (x-y)  \label{kcap3e45}
\end{equation}
\begin{equation}
\lbrack \hat{\Pi}_{i},\hat{\Pi}_{j}]=i\left( {\hat{\phi}_{j}(y)\hat{\Pi}%
_{i}(x)-\hat{\phi}_{i}(x)\hat{\Pi}_{j}(y)}\right)  \delta (x-y) .
\label{kcap3e46}
\end{equation}

\section{The non-minimal gauged O(3)
nonlinear sigma model} \label{sec:2} Now we construct a nonminimal
version of the gauged O(3) sigma model described by the Lagrangian
\begin{equation}
L=\frac{1}{2}\nabla _{\mu }{\bm \phi }\cdot \nabla ^{\mu }{\bm \phi }+%
\frac{\kappa }{2}\epsilon ^{\mu \nu \lambda }A_{\mu }\partial _{\nu
}A_{\lambda }-A_{\mu }\partial ^{\mu }b+\frac{\alpha }{2}b^{2},
\label{kcap3e47}
\end{equation}
where the minimal gauge covariant derivative was changed by
\begin{equation}
\nabla _{\mu }{\bm \phi =\partial }_{\mu }{\bm \phi +}\left[ eA_{\mu }+\frac{%
g}{2}\epsilon _{\mu \nu \sigma }F^{\nu \sigma }\right] {\bf n\times
\bm \phi } .\label{kcap3e48}
\end{equation}
Here $F^{\mu \nu}$ stands for the field-strength of the gauge field, the Levi-Civita symbol $%
\varepsilon _{\mu \nu \lambda }$ is fixed by $\varepsilon _{012}=1$
and $g$ is the nonminimal coupling constant.

Proceeding closely to the formulation for the minimal case we obtain
the canonical momenta conjugate to the fields \ $A_{0},$ $A_{k},$
$b,$ $\phi _{1},$ $\phi _{2},$ $\phi _{3}$ respectively, namely
\begin{equation}
\pi _{0}=0  \label{kcap3e49}
\end{equation}
\begin{equation}
\pi ^{k}=\frac{\kappa }{2}\epsilon ^{ki}A_{i}+g\epsilon ^{kj}{\bf
n}\times
{\bm \phi }\cdot D_{j}{\bm \phi +}g^{2}F^{0k}\left| {\bf n}\times {\bm \phi }%
\right| ^{2}  \label{kcap3e50}
\end{equation}
\begin{equation}
\pi _{b}=-A_{0}  \label{kcap3e51}
\end{equation}
\begin{equation}
\Pi _{1}=\dot{\phi}_{1}-eA_{0}\phi _{2}-\frac{g}{2}\epsilon
_{ij}F^{ij}\phi _{2}  \label{kcap3e52}
\end{equation}
\begin{equation}
\Pi _{2}=\dot{\phi}_{2}+eA_{0}\phi _{1}+\frac{g}{2}\epsilon
_{ij}F^{ij}\phi _{1}  \label{kcap3e53}
\end{equation}
\begin{equation}
\Pi _{3}=\partial _{0}\phi _{3}  .\label{kcap3e54}
\end{equation}

Note that $D_{j}{\bm \phi =\partial }_{j}{\bm \phi }+eA_{j}{\bf n}\times {\bf %
\phi }$. The primary constraints for the Lagrangian (\ref{kcap3e47})
are
\begin{equation}
V_{1}=\pi _{0}\approx 0  \label{kcap3e55}
\end{equation}
\begin{equation}
V_{2}={\bm \phi }\cdot {\bm \phi }-1\approx 0 \label{kcap3e56}
\end{equation}
\begin{equation}
V_{3}=\pi _{b}+A_{0}\approx 0  \label{kcap3e57}
\end{equation}
Therefore the canonical Hamiltonian can be written as
\begin{eqnarray}
H_{c} &=&\frac{1}{2}\frac{1}{g^{2}\left| {\bf n}\times {\bm \phi
}\right| ^{2}}[{\bf \pi }^{k}\pi _{k}+\kappa \epsilon _{jk}\pi
^{k}A_{j}+\frac{\kappa
^{2}}{4}A_{j}A^{j}+\kappa gA_{j}{\bf n}\times {\bm \phi }\cdot D^{j}{\bf %
\phi }  \nonumber \\
&&+g^{2}({\bf n}\times {\bm \phi }\cdot D^{j})({\bm \phi n}\times {\bm \phi }%
\cdot D_{j}{\bm \phi )+}2g\pi ^{k}\epsilon _{jk}{\bf n}\times {\bm \phi }%
\cdot D^{j}{\bm \phi ]}  \nonumber \\
&&+g\epsilon _{kj}\partial ^{k}A_{0}{\bf n}\times {\bm \phi }\cdot D^{j}{\bf %
\phi }+A_{j}\partial ^{j}b-\frac{\alpha
}{2}b^{2}-\frac{\kappa}{2}\epsilon ^{ij}A_{0}\partial
_{i}A_{j}+\frac{{\bf \Pi }\cdot {\bf \Pi }}{2}  \nonumber
\\
&&+\left( eA_{0}+\frac{g}{2}\epsilon _{ij}F^{ij}\right) \cdot \left(
\phi
_{2}\Pi _{1}-\phi _{1}\Pi _{2}\right) +\pi ^{k}\partial _{k}A_{0}-\frac{1}{2}%
D^{j}{\bm \phi }\cdot D_{j}{\bm \phi }  \nonumber \\
&&{\bf +}\frac{1}{4}g^{2}\left[ \frac{1}{2}\left[ \epsilon
_{ij}F^{ij}\right] ^{2}-F_{ij}F^{ij}\right] \left( \phi
_{1}^{2}+\phi _{2}^{2}\right) . \label{kcap3e58}
\end{eqnarray}

Now we address the issue of the existence of secondary and tertiary
constraints in the model. We note that the consistency conditions
$\dot{V}_{1}\approx 0$ e $\dot{V}_{3}\approx 0$ leads us to the
Lagrange multipliers $\lambda _{3}$ e $\lambda _{1}$ respectively.
On the other hand, the consistency condition $\dot{V}_{2}\approx 0$
give us a secondary constraint ($V_{4}$) and $\dot{V}_{4}\approx 0$
provide the Lagrange multiplier $\lambda _{1}$. There is no more
secondary (or tertiary) constraints, therefore our final set of
constraints is
\begin{equation}
V_{1}=\pi _{0}\approx 0  \label{kcap3e59}
\end{equation}
\begin{equation}
V_{2}={\bm \phi }\cdot {\bm \phi }-1\approx 0 \label{kcap3e60}
\end{equation}
\begin{equation}
V_{3}={\bf \pi }_{b}+A_{0}\approx 0  \label{kcap3e61}
\end{equation}
\begin{equation}
V_{4}={\bf \Pi }\cdot {\bm \phi }\approx 0  \label{kcap3e62}
\end{equation}
These constraints are clearly of second-class since
\begin{equation}
\lbrack V_{1},V_{3}]=-\delta (x-y)  \label{kcap3e63}
\end{equation}
\begin{equation}
\lbrack V_{2},V_{4}]=2{\bm \phi }({\bf x})\cdot {\bm \phi }({\bf
y})\delta (x-y)  .\label{kcap3e65}
\end{equation}
This is not unexpected since the Lagrangian (\ref{kcap3e47}) is
gauge fixed as in the previous case. It is worth mentioning that the
simpler structure of constraints (\ref{kcap3e59}-\ref{kcap3e62}) is
due to the presence of derivative terms in the canonical momenta
conjugate to the fields $A_{k}$, $\phi _{1}$, $\phi _{2}$,
represented by the Eqs. (\ref{kcap3e50}), (\ref{kcap3e52}) and
(\ref{kcap3e53}), respectively.

The matrix $C(x,y)$ in this case has the form
\begin{equation}
C(x,y)=\left[
\begin{array}{cccc}
0 & 0 & -1 & 0 \\
0 & 0 & 0 & 2 \\
1 & 0 & 0 & 0 \\
0 & -2 & 0 & 0
\end{array}
\right] \delta (x-y). \label{kcap3e66}
\end{equation}
Using the inverse of the matrix (\ref{kcap3e66}), we determine the
nonvanishing Dirac brackets
\begin{equation}
\lbrack A_{0},b]_{D}=\delta (x-y)  \label{kcap3e68}
\end{equation}
\begin{equation}
\lbrack A_{i},\pi _{j}]_{D}=\delta _{ij}\delta (x-y)
\label{kcap3e69}
\end{equation}
\begin{equation}
\lbrack \phi _{i},\Pi _{j}]_{D}=\left( \delta _{ij}-{\phi
_{i}(x)\phi _{j}(y)}\right) \delta (x-y) \label{kcap3e70}
\end{equation}
\begin{equation}
\lbrack \Pi _{i},\Pi _{j}]_{D}=\left( {\phi _{j}(y)\Pi _{i}(x)-\phi
_{i}(x)\Pi _{j}(y)}\right) \delta (x-y) \label{kcap3e71}
\end{equation}

Consequently, the quantized theory is obtained through the
commutators
\begin{equation}
\lbrack \hat{A}_{0},\hat{b}]=i\delta (x-y)  \label{kcap3e72}
\end{equation}
\begin{equation}
\lbrack \hat{A}_{i},\hat{\pi}_{j}]=i\delta _{ij}\delta (x-y)
\label{kcap3e73}
\end{equation}
\begin{equation}
\lbrack \hat{\phi}_{i},\hat{\Pi}_{j}]=i\left( \delta
_{ij}-{\hat{\phi} _{i}(x)\hat{\phi}_{j}(y)}\right) \delta (x-y)
\label{kcap3e74}
\end{equation}
\begin{equation}
\lbrack \hat{\Pi}_{i},\hat{\Pi}_{j}]=i\left({\hat{\phi}_{j}(y)\hat{\Pi}%
_{i}(x)-\hat{\phi}_{i}(x)\hat{\Pi}_{j}(y)} \right) \delta (x-y)
\label{kcap3e75}
\end{equation}

Here we would like to call attention to a comparison between the
commutator relations of two cases. Note that in the nonminimal case
the dependence of the Chern-Simons coefficient $k$ disappears.

\section{Conclusions} \label{sec:3}
In summary, we have first investigated some effects on the sympletic
structure of the gauged O(3) nonlinear sigma model with Chern-Simons
term due to introducing of a nonminimal Pauli coupling.  To begin
with, we consider the model with a minimal coupling and a covariant
gauge fixing and after we treat the nonminimal version. We have used
the Dirac quantization formalism in order to accomplish the
quantization of the models. We found that the sympletic structure
was changed by the introduction of the Pauli coupling. Indeed, the
reduction of the number of constraints made the quantization
procedure simpler. Furthermore the commutator relations of the
nonminimal case seems to indicate that the Chern-Simons term plays
no role in the quantized theory and consequently leads to a no
existence of fractional spin in the theory. Nevertheless, this issue
requires more studies. It is worthwhile to mentioning that results
in an Abelian Chern-Simons-Higgs model coupled non-minimally to
matter fields seems to show the presence of fractional spin
\cite{aug}, even in the absence of the Chern-Simons term in the
theory. Therefore, it would be interesting to investigating the
possibility of fractional spin and statistics in this model. As a
matter of fact, this has been accomplished by Lee {\it et al.}
\cite{lee} in a slightly different context, but in the minimal case.
So far, to the best of our knowledge, no study was accomplished for
the nonminimal case.

\section*{Acknowledgments}
This work was supported in part by Conselho Nacional de
Desenvolvimento Cient\'{\i }fico e Tecnol\'{o}gico-CNPq and
Funda\c{c}\~{a}o Cearense de Amparo \`{a} Pesquisa-FUNCAP.


\section*{References}

\vspace*{6pt}

\begin{thebibliography}{99}
\bibitem{zee} Wilczek F and Zee A, Phys. Rev. Lett. {\bf 51} (1983) 2250

\bibitem{lauglin} Kalmeyer V and Laughlin R, Phys. Rev. Lett. {\bf 59} (1987) 2095

\bibitem{polyakov} Dzyaloshinskii I, Polyakov A and Wiegmann P, Phys. Lett. {\bf A127} (1988) 112

\bibitem{larsen} Verbin Y, Madsen S and Larsen A, Phys. Rev. {\bf D67} (2003) 085019

\bibitem{pani} Panigrahi P K, Roy S, and Scherer W, Phys. Rev. Lett. {\bf 61} (1998) 2827

\bibitem{pani1} Panigrahi P, S. Roy and W. Scherer, Phys. Rev. {\bf D38} (1998) 3199

\bibitem{han} Han C, Phys. Rev. {\bf D47} (1993) 5521

\bibitem{bowick} Bowick M, Karabali D and Wjewardhana L, Nucl. Phys. {\bf B271} (1986) 417

\bibitem{marcon} Cavalcante F S, Cunha M S and Almeida C A S, Phys. Lett. {\bf B475 } (2000) 315

\bibitem{dirac} Dirac P A M,  {\textit Lectures in Quantum Mechanics},
(Yeshiva University Press, New York 1964)

\bibitem{kogan} Kogan I I, Phys. Lett. {\bf B262} (1991) 83

\bibitem{lat} Latinsky S and Sorokin D, Mod. Phys. Lett. {\bf A6} (1991) 3525

\bibitem{torres} Torres M, Phys. Rev. {\bf D46} (1992) R2295

\bibitem{georgelin} Georgelin Y and Wallet J, Mod. Phys. Lett. {\bf A7} (1992) 1149

\bibitem{aug} Nobre F A S and Almeida C A S,  Phys. Lett. {\bf B455}(1999) 213

\bibitem{lee}  Lee T, Rao C N and Viswanathan K S, Phys. Rev. {\bf D39}(1989) 2350

\end{thebibliography}
\end{document}